\documentclass[sigconf]{acmart}

\usepackage{epsfig}
\usepackage{graphicx}
\usepackage{amsmath}
\usepackage{helvet}  %Required
\usepackage{courier}  %Required
\usepackage{url}  %Required
\usepackage{xcolor}
\usepackage{multirow}
\usepackage{subcaption}
\usepackage{array}
\usepackage{hyperref}

\usepackage{color, colortbl}
\definecolor{mygray}{gray}{0.9}

\newcolumntype{P}[1]{>{\centering\arraybackslash}p{#1}}

\def\BibTeX{{\rm B\kern-.05em{\sc i\kern-.025em b}\kern-.08emT\kern-.1667em\lower.7ex\hbox{E}\kern-.125emX}}

\begin{document}
% \acmSubmissionID{153}
%
% The "title" command has an optional parameter, allowing the author to define a "short title" to be used in page headers.
% \title{A Novel Evaluation Framework for Image2Text Generation Driven by Large Language Models}
% \title{A Novel Evaluation Framework Driven by Large Language Models for Image2Text Generation}
\title{DeepEyeNet: Generating Medical Report for Retinal Images}

\author{Jia-Hong Huang}
\affiliation{%
  \institution{School: University of Amsterdam, Netherlands ; Contact: j.huang@uva.nl ; Ph.D. supervisor: Prof. Marcel Worring}}
% \email{j.huang@uva.nl}

%
% The abstract is a short summary of the work to be presented in the article.
\begin{abstract}
The increasing prevalence of retinal diseases poses a significant challenge to the healthcare system, as the demand for ophthalmologists surpasses the available workforce. This imbalance creates a bottleneck in diagnosis and treatment, potentially delaying critical care. Traditional methods of generating medical reports from retinal images rely on manual interpretation, which is time-consuming and prone to errors, further straining ophthalmologists' limited resources.
This thesis investigates the potential of Artificial Intelligence (AI) to automate medical report generation for retinal images. AI can quickly analyze large volumes of image data, identifying subtle patterns essential for accurate diagnosis. By automating this process, AI systems can greatly enhance the efficiency of retinal disease diagnosis, reducing doctors' workloads and enabling them to focus on more complex cases.
The proposed AI-based methods address key challenges in automated report generation: (1) A multi-modal deep learning approach captures interactions between textual keywords and retinal images, resulting in more comprehensive medical reports; (2) Improved methods for medical keyword representation enhance the system's ability to capture nuances in medical terminology; (3) Strategies to overcome RNN-based models' limitations, particularly in capturing long-range dependencies within medical descriptions; (4) Techniques to enhance the interpretability of the AI-based report generation system, fostering trust and acceptance in clinical practice.
These methods are rigorously evaluated using various metrics and achieve state-of-the-art performance. This thesis demonstrates AI's potential to revolutionize retinal disease diagnosis by automating medical report generation, ultimately improving clinical efficiency, diagnostic accuracy, and patient care.
\textbf{\href{https://github.com/Jhhuangkay/DeepOpht-Medical-Report-Generation-for-Retinal-Images-via-Deep-Models-and-Visual-Explanation}{DeepEyeNet Project Github.}}

\end{abstract}

\maketitle

\section{Introduction}

\begin{figure}[t!]
\begin{center}
\includegraphics[width=0.95\linewidth]{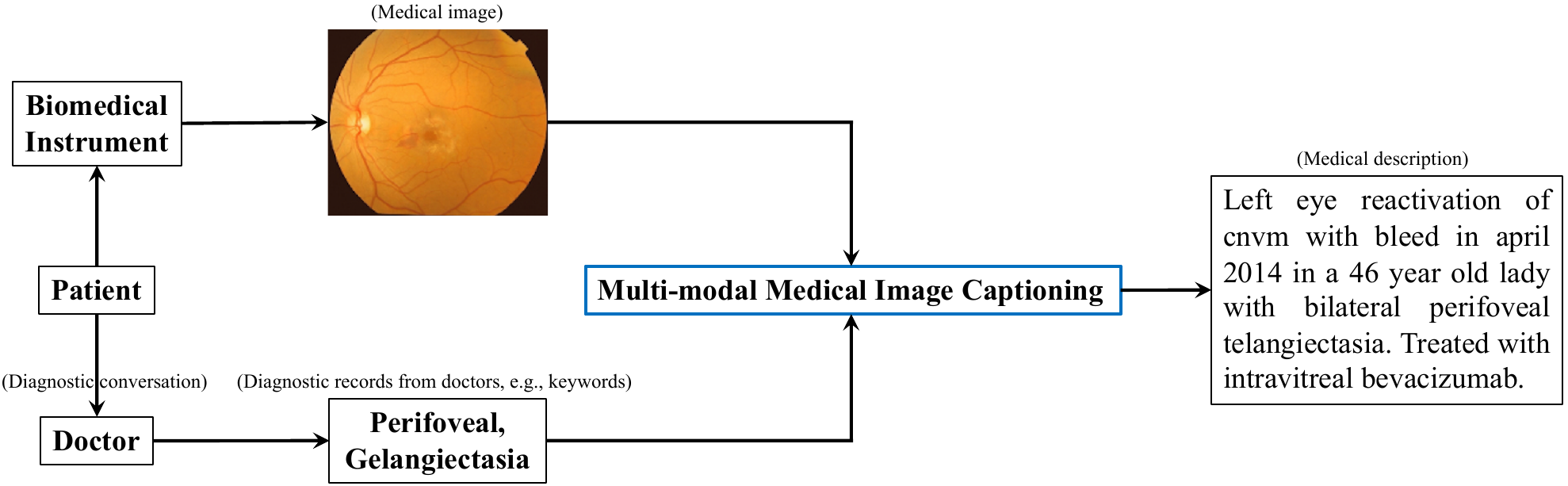}
\end{center}
\vspace{-0.5cm}
   \caption{Automated medical report generation. 
   A multi-modal medical report generation algorithm takes an image, e.g., a retinal image, and text-based diagnostic records, e.g., a set of keywords, as inputs to generate a medical description.
   }
\vspace{-0.4cm}
\label{fig:main}
\end{figure}

Medical imaging, particularly retinal imaging, plays a critical role in diagnosing and managing retinal diseases, capturing detailed visual information, providing context for patient conditions, representing temporal changes in disease progression, offering clinical realism, facilitating precise analysis, and enabling effective communication among healthcare providers. The increasing prevalence of retinal diseases and the advancements in imaging technologies have resulted in substantial growth of retinal image data generated daily in clinical settings \cite{yang2018novel,liu2019synthesizing,huck2019auto,huang2021deepopht}. While this abundance of medical imagery serves as a valuable source of information for diagnosing and treating retinal diseases, it also presents challenges for clinicians who need to efficiently and accurately interpret these images and generate medical reports to make informed treatment decisions.
In response to this challenge, automated medical report generation techniques have emerged, aiming to extract the most significant and relevant information from retinal images and present it in a concise, interpretable format, as depicted in Figure \ref{fig:main}.

The quality of automated medical reports is crucial for their utility in clinical practice and research. Inaccurate or incomplete reports can lead to misdiagnosis and suboptimal treatment, negatively impacting patient outcomes. In research, poorly generated reports can hinder study interpretations and slow medical advancements. Therefore, developing techniques that generate accurate, comprehensive, and relevant reports from retinal images, while minimizing irrelevant or misleading information, is essential for ensuring their reliability and effectiveness in both clinical and research contexts.

\begin{figure*}[t!]
\begin{center}
\includegraphics[width=0.95\linewidth]{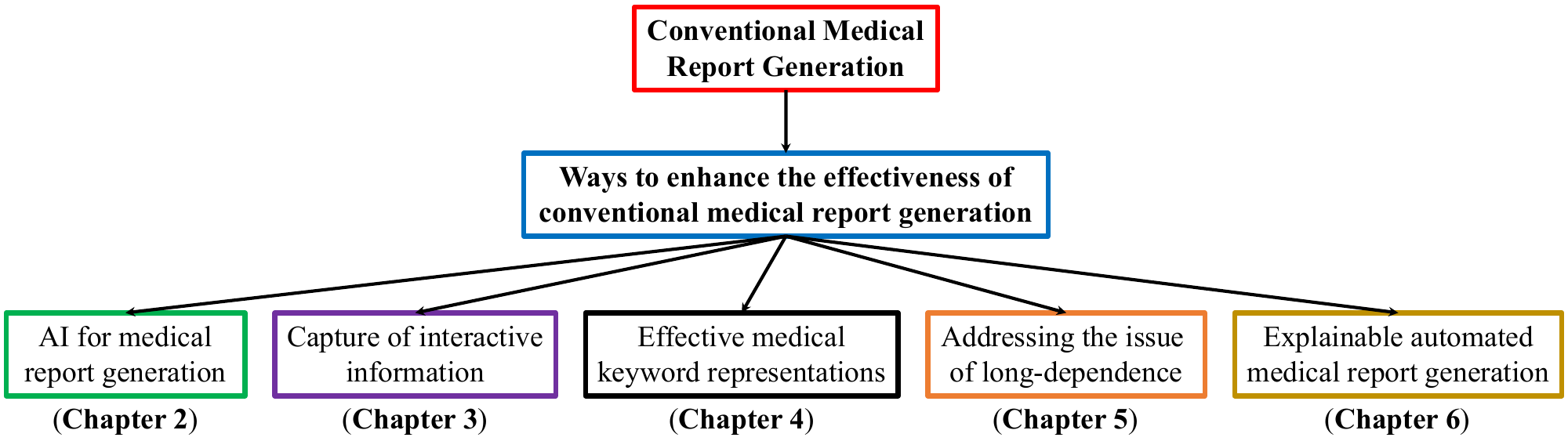}
\end{center}
\vspace{-0.5cm}
   \caption{An overview of the thesis.}
\vspace{-0.4cm}
\label{fig_bias_dl}
\end{figure*}

In this thesis, illustrated in Figure \ref{fig_bias_dl} for an overview, we therefore pose the central research question: \\
\noindent\textit{\textbf{How can we optimize the effectiveness and efficiency of the traditional treatment process for retinal diseases?}}

\begin{figure}[t!]
  \includegraphics[width=1.0 \linewidth]{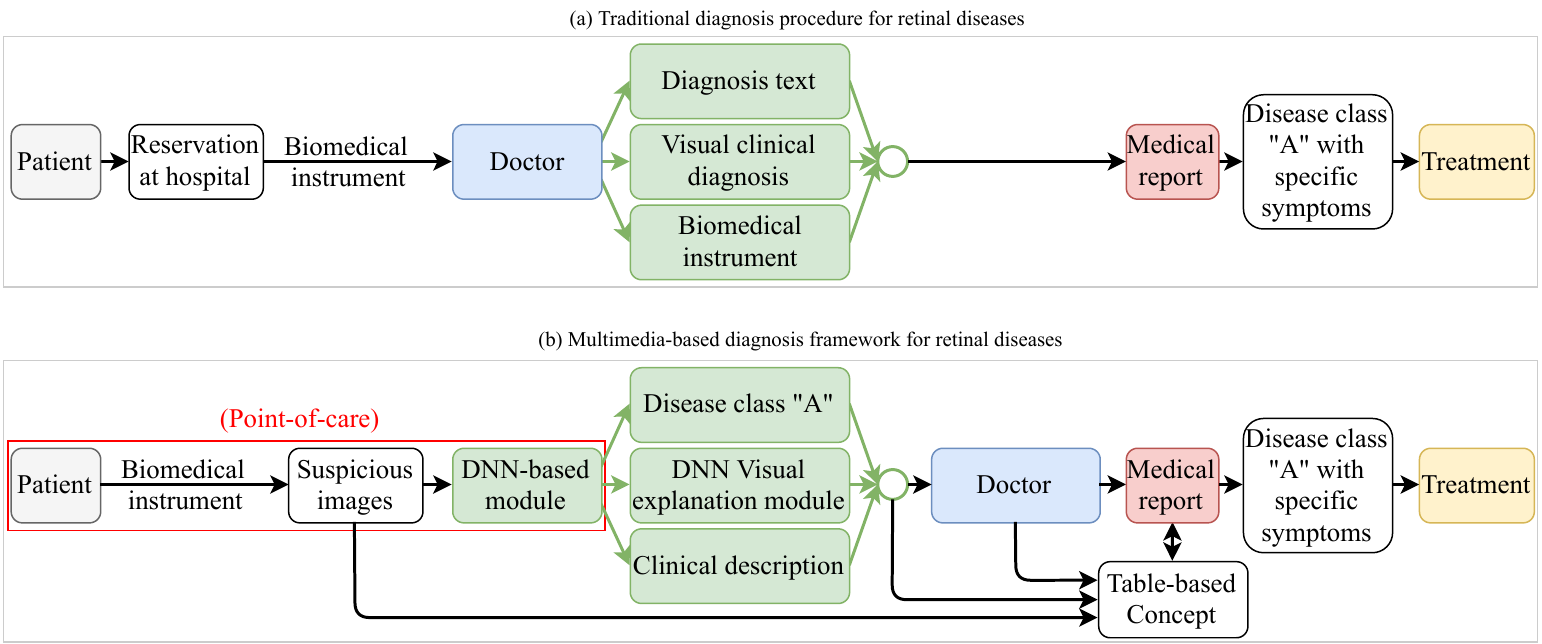}
    \vspace{-0.8cm}
  \caption{(a) is an existing traditional medical treatment process for retinal diseases \cite{tukey2014impact}. Typically, doctors have to handle most of the jobs in the traditional procedure. In (b), we incorporate the AI-based medical diagnosis method, referring to Figure \ref{fig:figure100}, in the conventional treatment procedure to improve the efficiency of (a).}
  \label{fig:flowchart_final_1}
  \vspace{-0.4cm}
\end{figure}

\begin{figure}[t!]
  \includegraphics[width=0.9 \linewidth]{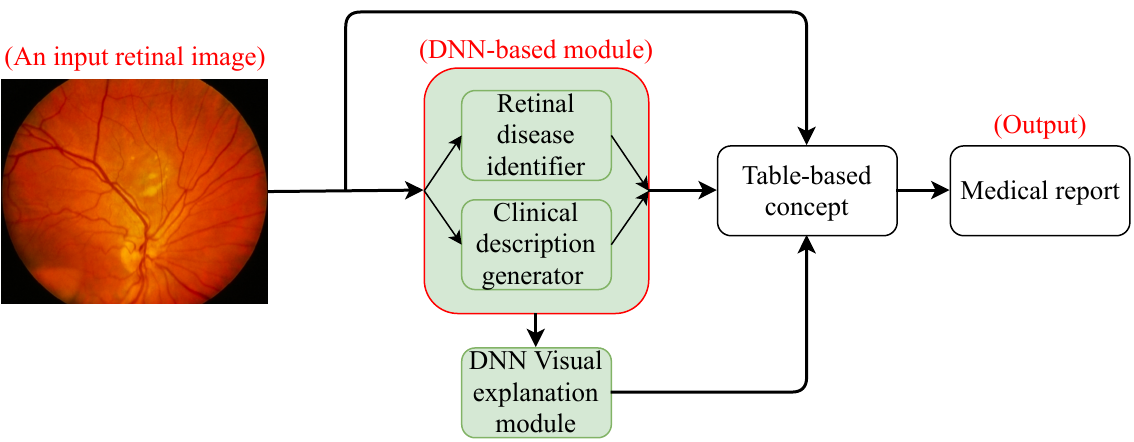}
  \vspace{-0.4cm}
  \caption{Our proposed AI-based medical diagnosis method in the ophthalmology expert domain \cite{huang2021deepopht}. The DNN-based module is composed of two sub-modules, i.e., a retinal disease identifier and a clinical description generator reinforced by our proposed keyword-driven method. In Figure \ref{fig:flowchart_final_1}, we show how to exploit the proposed AI-based method to improve the traditional retinal disease treatment procedure.}
%   \Description{Enjoying the baseball game from the third-base seats. Ichiro Suzuki preparing to bat.}
  \label{fig:figure100}
  \vspace{-0.4cm}
\end{figure}

Traditional methods for generating medical reports from retinal images rely on manual interpretation by clinicians, which is time-consuming and prone to human error (Figure \ref{fig:flowchart_final_1}). Reviewing numerous retinal images and patient records to create a comprehensive report is labor-intensive and subject to variability in expertise. Additionally, manual methods may struggle to consistently capture subtle nuances and complex patterns in retinal images, which are critical for accurate diagnosis. To address these limitations, we propose leveraging AI to automate parts of the report generation process. AI can quickly and accurately analyze vast amounts of image data, identifying patterns and anomalies that may be missed by humans. Integrating AI into the traditional diagnostic process aims to improve accuracy, reduce clinicians' workload, and enhance patient care.
This leads to the research question: \\
\noindent\textit{\textbf{How can we leverage AI to improve the existing retinal disease treatment procedure?}}
Traditional retinal disease diagnosis methods rely heavily on the manual interpretation of retinal images by ophthalmologists, which can be time-consuming and subject to human error \cite{jing-etal-2018-automatic,huck2019auto}. In Chapter 2 of this thesis, we investigate this issue and propose an AI-based method to improve the conventional retinal disease treatment procedure, building upon our previous work \cite{huang2021deepopht}.
As illustrated in Figure \ref{fig:figure100}, our approach consists of a deep neural networks-based (DNN-based) module, which includes a retinal disease identifier and a clinical description generator, along with a DNN visual explanation module. To train and validate the effectiveness of our DNN-based module, we have developed a large-scale retinal disease image dataset. Additionally, we provide a manually labeled retinal image dataset, annotated by ophthalmologists, to qualitatively demonstrate the effectiveness of the proposed AI-based method. Our method is capable of creating meaningful retinal image descriptions and visual explanations that are clinically relevant. The primary challenge lies in encoding and utilizing the interactive information between text-based keywords and retinal images to generate accurate and comprehensive medical reports. 
To address this, we present a multi-modal deep learning method capable of handling both image and text information, optimizing the generation of medical reports. We fuse these modalities using an average-based method to ensure the quality of the generated reports. Evaluating the model's performance is crucial, and we employ metrics such as BLEU, CIDEr, and ROUGE to assess the quality of the generated reports.  
Our experimental results show that the proposed method is effective both quantitatively and qualitatively, offering significant improvements in the diagnosis and treatment of retinal diseases.

\begin{figure}[t!]
  \includegraphics[width=0.9 \linewidth]{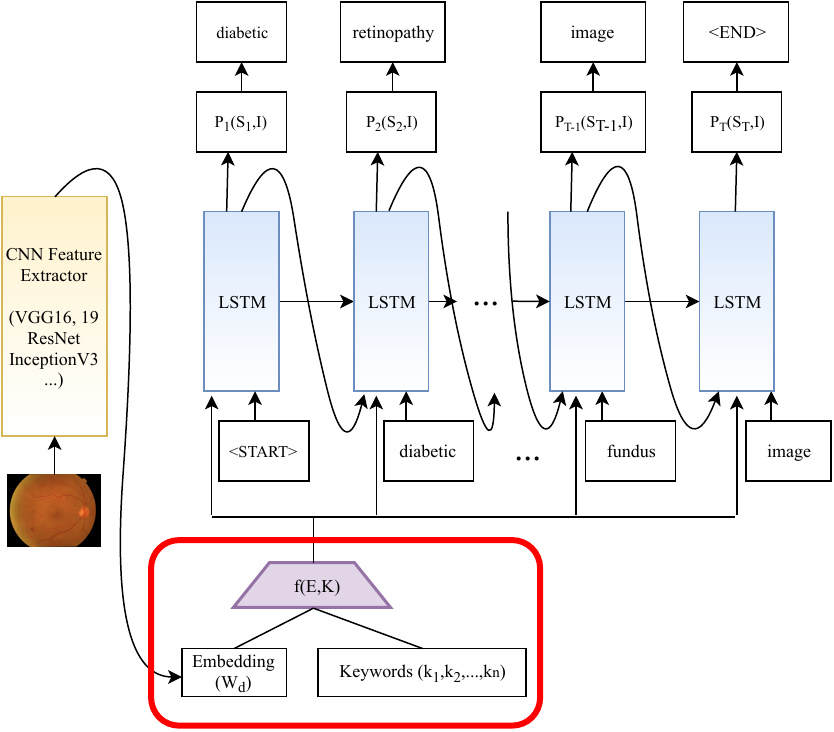}
    \vspace{-0.4cm}
  \caption{Our proposed context-driven network \cite{huang2021deep,huang2021longer}.}
  \label{fig:Basic-Keyword}
  \vspace{-0.4cm}
\end{figure}

In the proposed AI-based method, both text-based keywords and retinal images are integral to generating medical reports, with the interaction between these elements significantly influencing the quality of the reports. Therefore, it is crucial to develop effective approaches for encoding this interactive information. This prompts the next research question: \\
\noindent\textit{\textbf{How can we enhance the capture of interactive information between keywords and images in context-driven medical report generation?}}
Automatically generating medical reports for retinal images has emerged as a transformative approach to alleviate the workload of ophthalmologists and enhance clinical efficiency \cite{huck2019auto,huang2021deepopht,jing-etal-2018-automatic}. In Chapter 3, we introduce a novel context-driven network designed specifically for automating the generation of medical reports from retinal images, mainly built upon our earlier publications \cite{huang2021deep,huang2021longer}. Our proposed model integrates a multi-modal input encoder and a fused-feature decoder, aimed at effectively capturing the interactive information between textual keywords and retinal images, as illustrated in Figure \ref{fig:Basic-Keyword}. This architecture not only enhances the generation of accurate and meaningful descriptions but also addresses the nuanced complexities inherent in medical image interpretation. Our approach offers a robust framework for improving diagnostic accuracy and clinical decision-making in ophthalmology.

\begin{figure}[t!]
  \includegraphics[width=0.9 \linewidth]{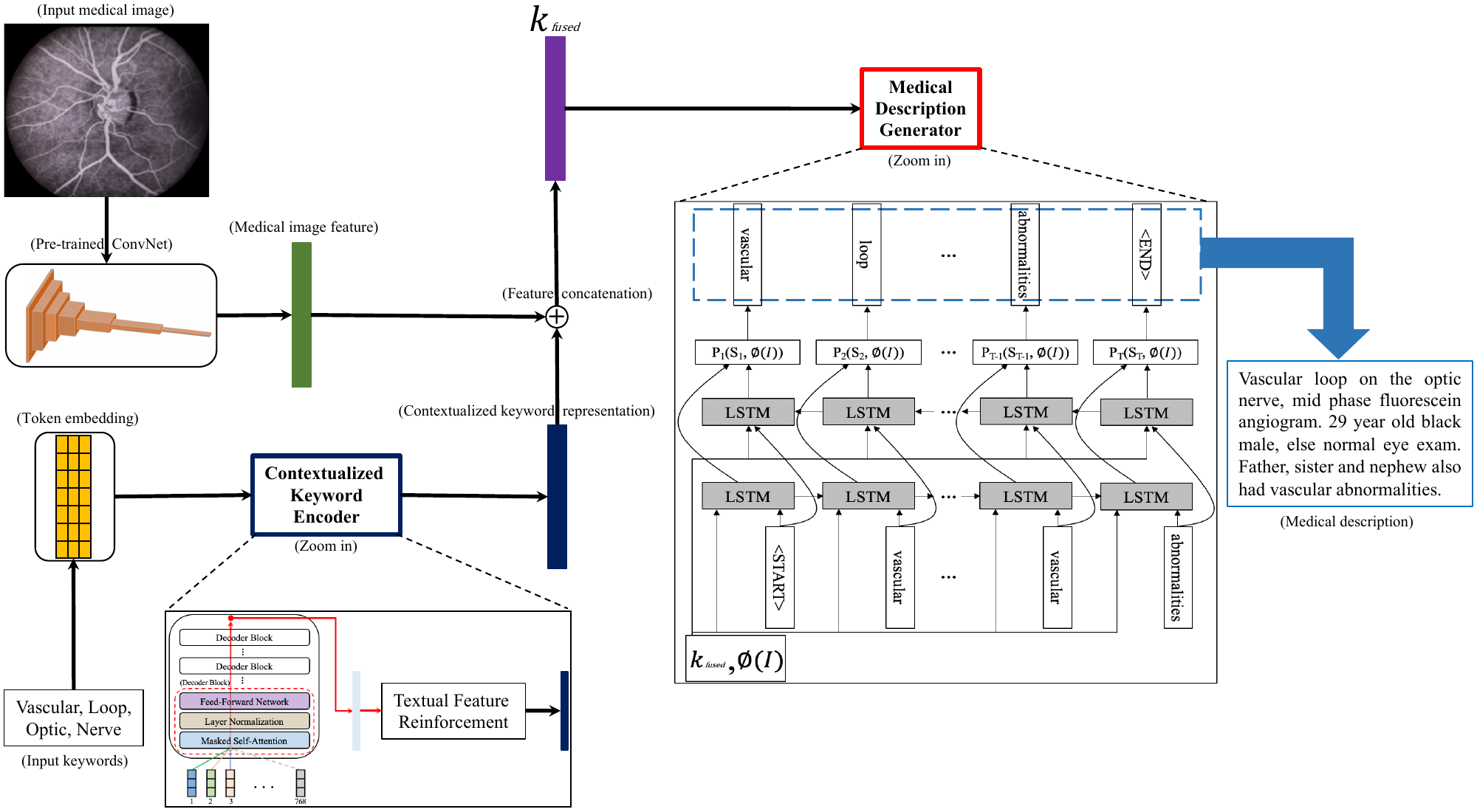}
    \vspace{-0.4cm}
  \caption{Our introduced contextualized keyword representations for multi-modal medical image captioning \cite{huang2021contextualized}. 
  % A pre-trained CNN, e.g., VGG16 or VGG19 pre-trained on ImageNet, is used to extract features from the medical image input (dark green). From a set of input keywords, the ``Token embedding'' generates the input to the ``Contextualized Keyword Encoder'' which is composed of a stack of decoder blocks and ``Textual Feature Reinforcement''. Each decoder block consists of the masked self-attention, layer normalization, and feed-forward network (red dashed line box). ``Textual Feature Reinforcement'', i.e., a stack of fully-connected layers, generates the contextualized keyword representation (dark blue). Note that $768$ color-coded brick-stacked vectors are the input of the keyword encoder. $\oplus$ indicates the concatenation of the medical image feature and contextualized keyword representation. In ``Medical Description Generator'' which creates the medical description, $k_{fused}$ denotes a fused feature vector (purple), $\phi(I)$ denotes an image feature vector, and $P_{i}(S_{i}, \phi(I))$ is a probability distribution where $i=1,2,...,T$.
  }
  \label{fig:flowchart_final}
  \vspace{-0.4cm}
\end{figure}

In multi-modal medical image captioning, the text-based input serves as critical guidance that significantly impacts the quality of the resulting reports. Hence, developing effective methods for encoding the provided text-based input is imperative. This prompts the next research question: \\
\noindent\textit{\textbf{How to improve the effectiveness of medical keyword representations to better capture the nuances of medical terminology?}}
Multi-modal medical image captioning, which combines visual and textual information, enhances the utility and effectiveness of automatically generated medical descriptions. By integrating diverse information sources, this approach creates comprehensive descriptions that capture both content and contextual nuances. In ophthalmology, combining expert-defined keywords with retinal images improves diagnostic accuracy and contextual relevance, enriching medical reports with detailed explanations. The success of multi-modal medical image captioning depends on proficiently encoding both textual input and medical images.

In Chapter 4, we introduce a novel end-to-end deep multi-modal medical image captioning model that leverages contextualized keyword representations, textual feature reinforcement, and masked self-attention. This chapter builds upon our earlier work cited in \cite{huang2021contextualized}. This model includes a contextualized medical description generator, multi-modal attention mechanisms, and a specially designed attention network, as shown in Figure \ref{fig:flowchart_final}. Unlike traditional single-image input methods, multi-modal inputs effectively capture semantic meaning and context, enhancing the overall quality of medical image captioning. Experimental results show that our model generates more accurate and meaningful descriptions for retinal images compared to baseline methods.

\begin{figure}[t!]
\begin{center}
\includegraphics[width=0.9\linewidth]{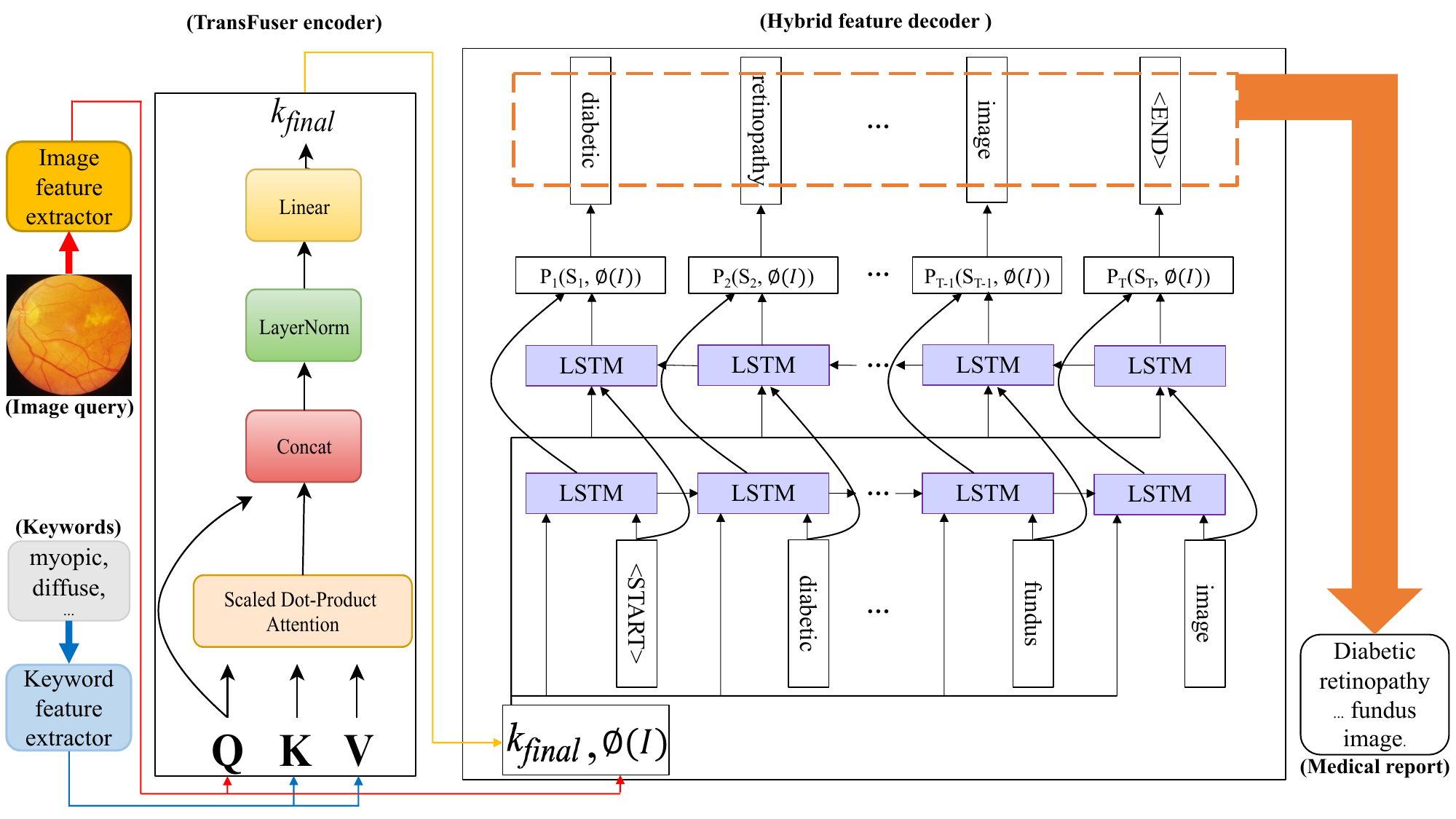}
\end{center}
\vspace{-0.5cm}
   \caption{Flowchart of our keyword-driven medical report generation method reinforced by our proposed multi-modal feature fuser TransFuser \cite{huang2022non}. It takes two inputs, a retinal image, and keywords. The purpose of keywords is to reinforce the model to generate more accurate and meaningful descriptions for retinal images.}
\vspace{-0.40cm}
\label{fig:figure777}
\end{figure}

In traditional automated medical report generation, RNN-based units are commonly employed for constructing report generation networks. However, due to the lengthy nature of medical reports, RNN-based units often struggle with long-dependency issues, which are well-documented. These issues can significantly degrade the quality of generated medical reports. Hence, this leads to the following research question: \\
\noindent\textit{\textbf{How can we address the long-dependency issue inherent in RNN-based models for generating medical descriptions?}}
In addressing the long-dependency issue inherent in RNN-based models for generating medical descriptions from retinal images, Chapter 5, built upon our prior work \cite{huang2022non}, introduces a novel approach, inspired by how ophthalmologists utilize expert-defined keywords to capture crucial information early in diagnosis. Existing methods, relying solely on image data, often struggle with abstract medical concepts and long-dependency issues. To address this, we integrate additional information through expert-defined keywords, enhancing the effectiveness of report generation.

Our approach utilizes TransFuser, a novel non-local attention-based multi-modal feature fusion technique (see Figure \ref{fig:figure777}). TransFuser effectively merges features from both retinal images and unordered keywords, resolving the long-dependency issue. By capturing the essential mutual information between these elements, TransFuser enables the generation of more precise and informative medical descriptions. Experimental results demonstrate that our keyword-driven model, empowered by TransFuser, significantly outperforms baseline methods on established text evaluation metrics like BLEU, CIDEr, and ROUGE.
% Beyond improving performance, this approach enhances the interpretability of the report generation process. By systematically integrating keywords with image data, the method provides clearer insights into how different inputs contribute to the final medical report. This chapter illustrates that our keyword-driven generation model achieves state-of-the-art performance.

\begin{figure}[t!]
  \includegraphics[width=0.9 \linewidth]{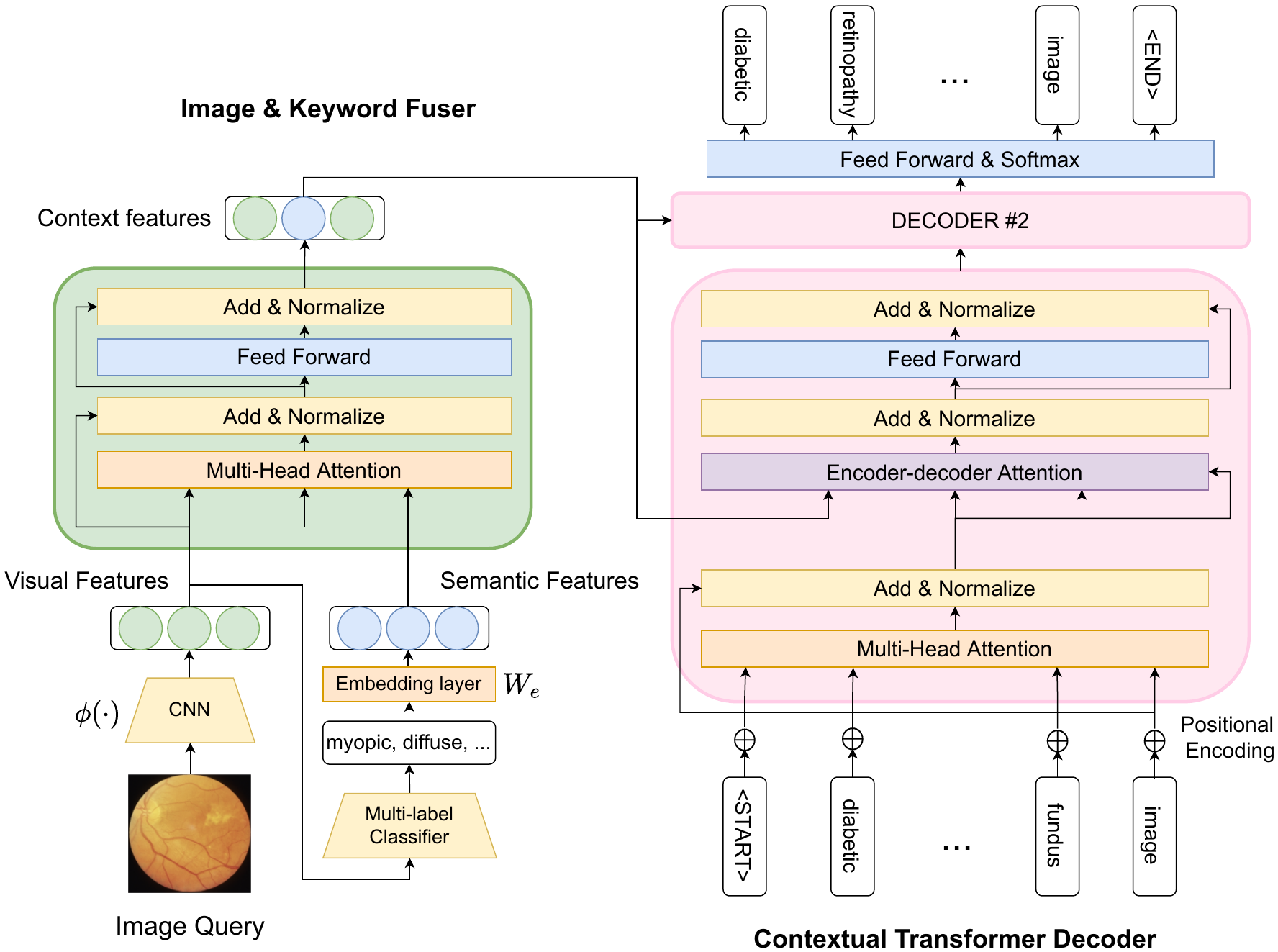}
    \vspace{-0.4cm}
  \caption{Flowchart of the proposed method \cite{wu2023expert}.}
  \label{fig:figure_222}
  \vspace{-0.6cm}
\end{figure}

Existing automated medical report generation methods typically rely on deep learning. However, the lack of explainability in deep models can reduce user trust, particularly in the medical domain. Consequently, the final research question emerges: \\
\noindent\textit{\textbf{How can we improve the explainability of automated medical report generation for retinal images?}}
The interpretability of machine learning (ML)-based medical report generation systems for retinal images remains a significant challenge, hindering their widespread acceptance. Chapter 6 addresses this crucial issue, emphasizing the need for trust in ML-based systems. Building on our previous work \cite{wu2023expert}, this chapter explores the complexities of defining interpretability in a way that makes ML-based medical report generation systems human-comprehensible.
Common post-hoc explanation methods, like heat maps and saliency maps, highlight important image areas but fail to convey the specific features the model finds useful. This discrepancy can lead to biases due to varying human interpretations of these highlighted regions.

To tackle these issues, we propose using expert-defined keywords and a specialized attention-based strategy to enhance interpretability in medical report generation systems for retinal images (see Figure \ref{fig:figure_222}). These keywords are effective carriers of domain knowledge and are inherently understandable, improving the system's interpretability. By integrating these keywords with an attention-based mechanism, our approach enhances both interpretability and performance.
Our method achieves state-of-the-art performance on text evaluation metrics such as BLEU, ROUGE, CIDEr, and METEOR. This chapter demonstrates that incorporating expert-defined keywords and a robust attention-based strategy can create a more trustworthy and interpretable ML-based medical report generation system, overcoming a critical barrier to clinical acceptance.

%%%%%%%%%%%%%%%%%%%%%%%%%%%%%%%%%%%%%%%%%%%%%%%%%
%%%%%%%%%%%%      Methodology       %%%%%%%%%%%%%
%%%%%%%%%%%%%%%%%%%%%%%%%%%%%%%%%%%%%%%%%%%%%%%%%

% \vspace{-0.1cm}
\section{Methods, Experimental Results, Conclusions and Future Work}
This thesis improves the traditional retinal disease treatment process by integrating text-based medical keywords with effective feature embeddings, effectively capturing interactive information between keywords and images, and enhancing explainability through specially designed attention visualizations with keywords.
Refer to Figures \ref{fig:figure100}, \ref{fig:Basic-Keyword}, \ref{fig:flowchart_final}, \ref{fig:figure777}, and \ref{fig:figure_222} for detailed explanations of our proposed methodology. The proposed experiments utilize commonly used retinal datasets: EyeNet \cite{yang2018novel} and DeepEyeNet \cite{huang2021deepopht}. Please refer to Table \ref{table:main1}, \ref{table:main2}, and \ref{table:correctness} for the results. Our proposed methods are effective. I plan to explore LLMs and diffusion models for further improvement in the retinal disease diagnosis process. I'm eager to discuss these future directions with mentors and peers at the symposium. See the appendix for more details of conclusions and future work.

\appendix
\section{Related Publications during Ph.D.}
\cite{huang2017robustnessMS,huang2019novel,huang2017vqabq,huang2018robustness,huang2023improving,huang2019assessing,huang2021deepopht,huang2022non,huang2021contextualized,huang2021deep,huang2021longer,wu2023expert,huck2019auto,liu2019synthesizing,yang2018novel,di2022dawn,huang2020query,huang2021gpt2mvs,huang2022causal,huang2023causalainer,huang2023query,huang2023conditional,hu2019silco,wang2024ada,zhu2024enhancing,huang2024multi,huang2024optimizing,huang2025image2text2image,huang2024novel,huang2024personalized,huang2024automated,zhang2024comparative,zhang2024beyond,zhang2024towards,zhang2024qfmts,shen2024parameterefficient,shen2025macp,shen2025ssh,huang2025gradient}.

\begin{table}[t!]
\caption{This table shows the evaluation results of the proposed model compared with several competitive baselines by using expert-defined keywords, i.e., ground truth keywords. ``BLEU-avg'' denotes the average score of BLEU-1, BLEU-2, BLEU-3, and BLEU-4. All the keyword-driven models are superior to the non-keyword-driven models.}
\vspace{-0.4cm}
\centering
\small
\scalebox{0.6}{
\begin{tabular}{|c||c|c|c|c|c|c|c|c|}
\hline
\textbf{Model}                                    & \textbf{BLEU-1} & \textbf{BLEU-2} & \textbf{BLEU-3} & \textbf{BLEU-4} & \textbf{B-avg} & \textbf{ROUGE} & \textbf{CIDEr} & \textbf{METEOR} \\ \hline\hline
LSTM \cite{wang2016image}                & 0.2273	& 0.1650	& 0.1224	& 0.1017	& 0.1541 & 0.2533	& 0.1102	& 0.2437 \\ \hline
Show and tell \cite{xu2016show}          & 0.4234	& 0.3583	& 0.3002	& 0.2757	& 0.3394 & 0.4463	& 0.3029	& 0.4335 \\ \hline\hline
Semantic Att \cite{you2016image}         & 0.5904	& 0.5100	& 0.4360	& 0.3969	& 0.4833 & 0.6228	& 0.4460	& 0.6056 \\
\hline
ContexGPT \cite{huang2021contextualized}  & 0.6254  & 0.5500    & 0.4758    & 0.4344    & 0.5214 & 0.6602   & 0.4951    & 0.6390 \\
\hline
% DeepContex \cite{huang2021deep}           & 0.6749  & 0.6036    & 0.5307    & 0.4890    & 0.5745 & 0.7020   & 0.5496    & 0.6835 \\
% \hline
CoAtt \cite{jing-etal-2018-automatic}     & 0.6712	& 0.5950	& 0.5211	& 0.4817	& 0.5673 & 0.6988	& 0.5419	& 0.6798 \\ \hline
H-CoAtt \cite{lu2017hierarchical}         & 0.6718	& 0.5956	& 0.5201	& 0.4829	& 0.5676 & 0.7045	& 0.5417	& 0.6864 \\ \hline
DeepContex \cite{huang2021deep}           & 0.6749  & 0.6036    & 0.5307    & 0.4890    & 0.5745 & 0.7020   & 0.5496    & 0.6835 \\
\hline
MIA \cite{liu2019aligning}                & 0.6877	& 0.6138	& 0.5421	& 0.5000	& 0.5859 & 0.7195	& 0.5596	& 0.7006 \\ \hline\hline
\rowcolor{mygray} Ours                              & \textbf{0.6969}	& \textbf{0.6195}	& \textbf{0.5496}	& \textbf{0.5008}	& \textbf{0.5892} & \textbf{0.7252}	& \textbf{0.5650}	& \textbf{0.7044} \\ \hline
\end{tabular}}
\label{table:main1}
\vspace{-0.4cm}
\end{table}

\begin{table}[t!]
\caption{This table shows the evaluation results of the proposed model compared with several competitive baselines by using predicted keywords, i.e., pseudo expert-defined keywords. ``BLEU-avg'' denotes the average score of BLEU-1, BLEU-2, BLEU-3, and BLEU-4. All the keyword-driven models are superior to the non-keyword-driven models.}
\vspace{-0.4cm}
\centering
\small
\scalebox{0.6}{
\begin{tabular}{|c||c|c|c|c|c|c|c|c|}
\hline
\textbf{Model}                                    & \textbf{BLEU-1} & \textbf{BLEU-2} & \textbf{BLEU-3} & \textbf{BLEU-4} & \textbf{B-avg} & \textbf{ROUGE} & \textbf{CIDEr} & \textbf{METEOR} \\ \hline\hline
LSTM \cite{wang2016image}           & 0.2273	& 0.1650	& 0.1224	& 0.1017	& 0.1541	& 0.2533	& 0.1102	& 0.2437 \\ \hline
Show and tell \cite{xu2016show}     & 0.4234	& 0.3583	& 0.3002	& 0.2757	& 0.3394	& 0.4463	& 0.3029	& 0.4335 \\ \hline\hline
H-CoAtt \cite{lu2017hierarchical}    & 0.4465	& 0.3822	& 0.3285	& 0.2969	& 0.3636	& 0.4788	& 0.3409	& 0.4564 \\
\hline
ContexGPT \cite{huang2021contextualized} & 0.4493 & 0.3744  & 0.3109    & 0.2800	& 0.3536    & 0.4771    & 0.3171    & 0.4588 \\ 
\hline
Semantic Att \cite{you2016image}     & 0.4541	& 0.3771	& 0.3117	& 0.2777	& 0.3552	& 0.4785	& 0.3118	& 0.4610 \\ \hline
CoAtt \cite{jing-etal-2018-automatic}& 0.4647	& 0.4038	& 0.3479	& 0.3162	& 0.3831	& 0.4906	& 0.3563	& 0.4759 \\ 
\hline
DeepContex \cite{huang2021deep}      & 0.4683   & 0.3966    & 0.3302    & 0.2969    & 0.3730    & 0.4941    & 0.3341    & 0.4803 \\
\hline
MIA \cite{liu2019aligning}           & 0.5077	& 0.4446	& 0.3861	& 0.3514	& 0.4224	& 0.5326	& 0.3897	& 0.5163 \\ \hline\hline
\rowcolor{mygray} Ours                         & \textbf{0.5268}	& \textbf{0.4600}	& \textbf{0.3915}	& \textbf{0.3634}	& \textbf{0.4354}	& \textbf{0.5482}	& \textbf{0.4105}	& \textbf{0.5316} \\ \hline
\end{tabular}}
\vspace{-0.4cm}
\label{table:main2}
\end{table}

\begin{table}[t!]
\caption{The table is to show the performance drop when expert-defined keywords are not available, i.e., the case ``With predicted keywords''.}
\vspace{-0.4cm}
\small
\centering
\scalebox{0.55}{
\begin{tabular}{|c||c|c|c|c|c|c|c|c|}
\hline
\textbf{Model}                                    & \textbf{BLEU-1} & \textbf{BLEU-2} & \textbf{BLEU-3} & \textbf{BLEU-4} & \textbf{B-avg} & \textbf{ROUGE} & \textbf{CIDEr} & \textbf{METEOR} \\ \hline\hline
With predicted keywords                          & 0.5268 & 0.4600 & 0.3915 & 0.3634 & 0.4354 & 0.5482 & 0.4105 & 0.5316 \\ \hline
\rowcolor{mygray} With expert-defined keywords                       & \textbf{0.6969}	& \textbf{0.6195}	& \textbf{0.5496}	& \textbf{0.5008}	& \textbf{0.5892} & \textbf{0.7252}	& \textbf{0.5650}	& \textbf{0.7044} \\ \hline
\end{tabular}}
\label{table:correctness}
\vspace{-0.4cm}
\end{table}

\begin{figure}[t!]
\begin{center}
\includegraphics[width=0.9\linewidth]{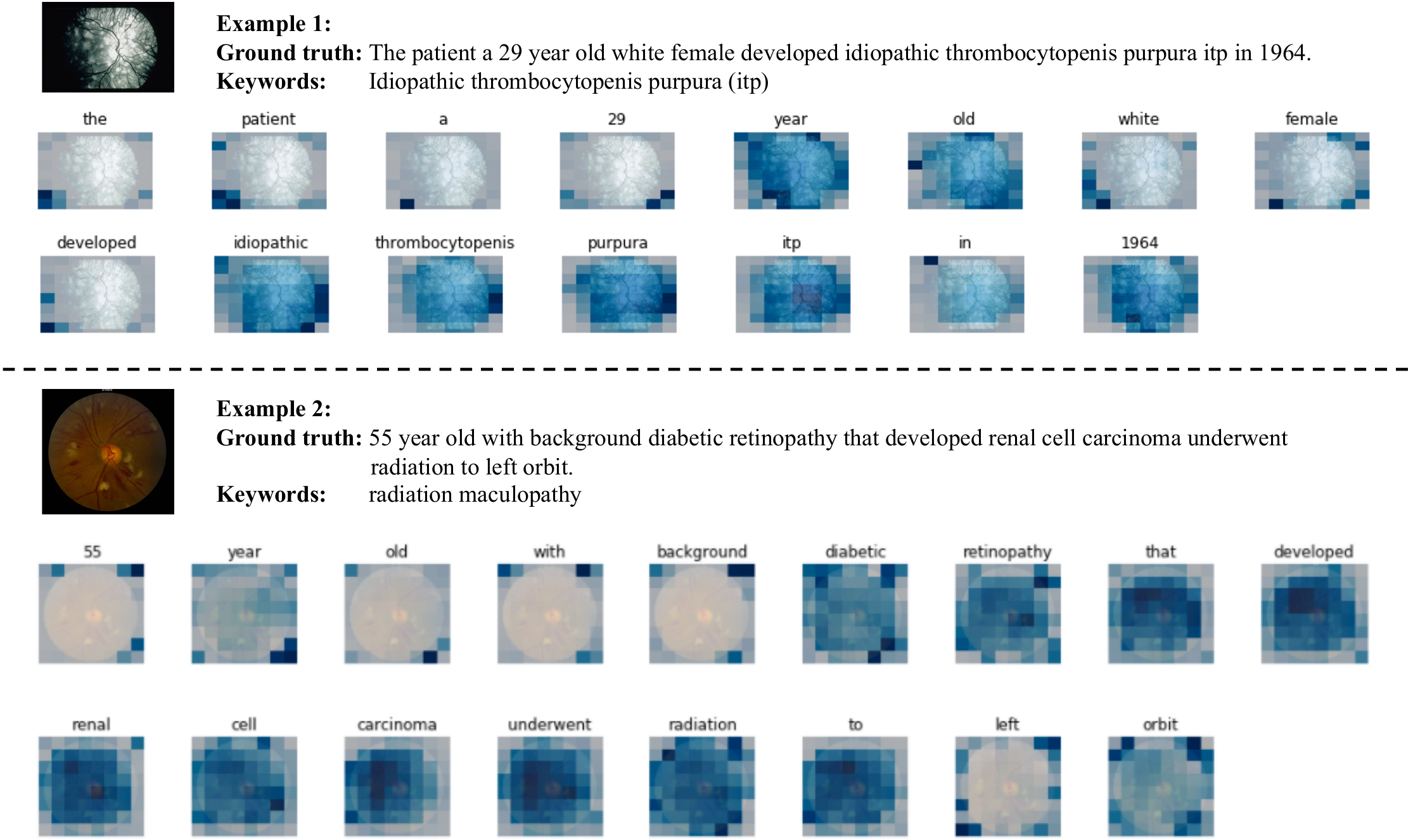}
\end{center}
\vspace{-0.5cm}
  \caption{The visualization shows image attention during text generation. Each example includes an input image, ground truth descriptions, and predicted keywords. Predicted words are displayed over the image attention at each time step, enhancing explainability.}
\vspace{-0.4cm}
\label{fig:vis2}
\end{figure}

% \begin{acks}
% This project has received funding from the European Union’s Horizon 2020 research and innovation programme under the Marie Skłodowska-Curie grant agreement No 765140.
% \end{acks}

%
% The next two lines define the bibliography style to be used, and the bibliography file.
\bibliographystyle{ACM-Reference-Format}
\bibliography{sample-base}


\begin{thebibliography}{46}

%%% the Web address.

\ifx \showCODEN    \undefined \def \showCODEN     #1{\unskip}     \fi
\ifx \showDOI      \undefined \def \showDOI       #1{#1}\fi
\ifx \showISBNx    \undefined \def \showISBNx     #1{\unskip}     \fi
\ifx \showISBNxiii \undefined \def \showISBNxiii  #1{\unskip}     \fi
\ifx \showISSN     \undefined \def \showISSN      #1{\unskip}     \fi
\ifx \showLCCN     \undefined \def \showLCCN      #1{\unskip}     \fi
\ifx \shownote     \undefined \def \shownote      #1{#1}          \fi
\ifx \showarticletitle \undefined \def \showarticletitle #1{#1}   \fi
\ifx \showURL      \undefined \def \showURL       {\relax}        \fi
% invisible to TeX
\providecommand\bibfield[2]{#2}
\providecommand\bibinfo[2]{#2}
\providecommand\natexlab[1]{#1}
\providecommand\showeprint[2][]{arXiv:#2}

\bibitem[\protect\citeauthoryear{Di~Sipio, Huang, Chen, Mangini, and Worring}{Di~Sipio et~al\mbox{.}}{2022}]%
        {di2022dawn}
\bibfield{author}{\bibinfo{person}{Riccardo Di~Sipio}, \bibinfo{person}{Jia-Hong Huang}, \bibinfo{person}{Samuel Yen-Chi Chen}, \bibinfo{person}{Stefano Mangini}, {and} \bibinfo{person}{Marcel Worring}.} \bibinfo{year}{2022}\natexlab{}.
\newblock \showarticletitle{The dawn of quantum natural language processing}. In \bibinfo{booktitle}{\emph{ICASSP 2022-2022 IEEE International Conference on Acoustics, Speech and Signal Processing (ICASSP)}}. IEEE, \bibinfo{pages}{8612--8616}.
\newblock


\bibitem[\protect\citeauthoryear{Hu, Mettes, Huang, and Snoek}{Hu et~al\mbox{.}}{2019}]%
        {hu2019silco}
\bibfield{author}{\bibinfo{person}{Tao Hu}, \bibinfo{person}{Pascal Mettes}, \bibinfo{person}{Jia-Hong Huang}, {and} \bibinfo{person}{Cees~GM Snoek}.} \bibinfo{year}{2019}\natexlab{}.
\newblock \showarticletitle{SILCO: Show a Few Images, Localize the Common Object}. In \bibinfo{booktitle}{\emph{Proceedings of the IEEE International Conference on Computer Vision}}. \bibinfo{pages}{5067--5076}.
\newblock


\bibitem[\protect\citeauthoryear{Huang}{Huang}{2017}]%
        {huang2017robustnessMS}
\bibfield{author}{\bibinfo{person}{Jia-Hong Huang}.} \bibinfo{year}{2017}\natexlab{}.
\newblock \showarticletitle{Robustness Analysis of Visual Question Answering Models by Basic Questions}.
\newblock \bibinfo{journal}{\emph{King Abdullah University of Science and Technology, Master Thesis}} (\bibinfo{year}{2017}).
\newblock


\bibitem[\protect\citeauthoryear{Huang}{Huang}{2024a}]%
        {huang2024automated}
\bibfield{author}{\bibinfo{person}{Jia-Hong Huang}.} \bibinfo{year}{2024}\natexlab{a}.
\newblock \showarticletitle{Automated Retinal Image Analysis and Medical Report Generation through Deep Learning}.
\newblock \bibinfo{journal}{\emph{University of Amsterdam, Doctoral Thesis}} (\bibinfo{year}{2024}).
\newblock


\bibitem[\protect\citeauthoryear{Huang}{Huang}{2024b}]%
        {huang2024multi}
\bibfield{author}{\bibinfo{person}{Jia-Hong Huang}.} \bibinfo{year}{2024}\natexlab{b}.
\newblock \showarticletitle{Multi-modal Video Summarization}. In \bibinfo{booktitle}{\emph{Proceedings of the 2024 International Conference on Multimedia Retrieval}}. \bibinfo{pages}{1214--1218}.
\newblock


\bibitem[\protect\citeauthoryear{Huang}{Huang}{2024c}]%
        {huang2024personalized}
\bibfield{author}{\bibinfo{person}{Jia-Hong Huang}.} \bibinfo{year}{2024}\natexlab{c}.
\newblock \showarticletitle{Personalized Video Summarization using Text-Based Queries and Conditional Modeling}.
\newblock \bibinfo{journal}{\emph{University of Amsterdam, Doctoral Thesis}} (\bibinfo{year}{2024}).
\newblock


\bibitem[\protect\citeauthoryear{Huang, Alfadly, and Ghanem}{Huang et~al\mbox{.}}{2017}]%
        {huang2017vqabq}
\bibfield{author}{\bibinfo{person}{Jia-Hong Huang}, \bibinfo{person}{Modar Alfadly}, {and} \bibinfo{person}{Bernard Ghanem}.} \bibinfo{year}{2017}\natexlab{}.
\newblock \showarticletitle{Vqabq: Visual question answering by basic questions}.
\newblock \bibinfo{journal}{\emph{IEEE/CVF Conference on Computer Vision and Pattern Recognition Visual Question Answering (VQA) Challenge Workshop}} (\bibinfo{year}{2017}).
\newblock


\bibitem[\protect\citeauthoryear{Huang, Alfadly, Ghanem, and Worring}{Huang et~al\mbox{.}}{2019a}]%
        {huang2019assessing}
\bibfield{author}{\bibinfo{person}{Jia-Hong Huang}, \bibinfo{person}{Modar Alfadly}, \bibinfo{person}{Bernard Ghanem}, {and} \bibinfo{person}{Marcel Worring}.} \bibinfo{year}{2019}\natexlab{a}.
\newblock \showarticletitle{Assessing the Robustness of Visual Question Answering}.
\newblock \bibinfo{journal}{\emph{arXiv preprint arXiv:1912.01452}} (\bibinfo{year}{2019}).
\newblock


\bibitem[\protect\citeauthoryear{Huang, Alfadly, Ghanem, and Worring}{Huang et~al\mbox{.}}{2023a}]%
        {huang2023improving}
\bibfield{author}{\bibinfo{person}{Jia-Hong Huang}, \bibinfo{person}{Modar Alfadly}, \bibinfo{person}{Bernard Ghanem}, {and} \bibinfo{person}{Marcel Worring}.} \bibinfo{year}{2023}\natexlab{a}.
\newblock \showarticletitle{Improving Visual Question Answering Models through Robustness Analysis and In-Context Learning with a Chain of Basic Questions}.
\newblock \bibinfo{journal}{\emph{arXiv preprint arXiv:2304.03147}} (\bibinfo{year}{2023}).
\newblock


\bibitem[\protect\citeauthoryear{Huang, Dao, Alfadly, and Ghanem}{Huang et~al\mbox{.}}{2019b}]%
        {huang2019novel}
\bibfield{author}{\bibinfo{person}{Jia-Hong Huang}, \bibinfo{person}{Cuong~Duc Dao}, \bibinfo{person}{Modar Alfadly}, {and} \bibinfo{person}{Bernard Ghanem}.} \bibinfo{year}{2019}\natexlab{b}.
\newblock \showarticletitle{A novel framework for robustness analysis of visual qa models}. In \bibinfo{booktitle}{\emph{Proceedings of the AAAI Conference on Artificial Intelligence}}, Vol.~\bibinfo{volume}{33}. \bibinfo{pages}{8449--8456}.
\newblock


\bibitem[\protect\citeauthoryear{Huang, Dao, Alfadly, Yang, and Ghanem}{Huang et~al\mbox{.}}{2018}]%
        {huang2018robustness}
\bibfield{author}{\bibinfo{person}{Jia-Hong Huang}, \bibinfo{person}{Cuong~Duc Dao}, \bibinfo{person}{Modar Alfadly}, \bibinfo{person}{C~Huck Yang}, {and} \bibinfo{person}{Bernard Ghanem}.} \bibinfo{year}{2018}\natexlab{}.
\newblock \showarticletitle{Robustness analysis of visual qa models by basic questions}.
\newblock \bibinfo{journal}{\emph{IEEE/CVF Conference on Computer Vision and Pattern Recognition Visual Question Answering Visual Question Answering (VQA) Challenge and Visual Dialog Workshop}} (\bibinfo{year}{2018}).
\newblock


\bibitem[\protect\citeauthoryear{Huang, Murn, Mrak, and Worring}{Huang et~al\mbox{.}}{2021a}]%
        {huang2021gpt2mvs}
\bibfield{author}{\bibinfo{person}{Jia-Hong Huang}, \bibinfo{person}{Luka Murn}, \bibinfo{person}{Marta Mrak}, {and} \bibinfo{person}{Marcel Worring}.} \bibinfo{year}{2021}\natexlab{a}.
\newblock \showarticletitle{GPT2MVS: Generative Pre-trained Transformer-2 for Multi-modal Video Summarization}. In \bibinfo{booktitle}{\emph{Proceedings of the International Conference on Multimedia Retrieval}}. \bibinfo{pages}{580--589}.
\newblock


\bibitem[\protect\citeauthoryear{Huang, Murn, Mrak, and Worring}{Huang et~al\mbox{.}}{2023b}]%
        {huang2023query}
\bibfield{author}{\bibinfo{person}{Jia-Hong Huang}, \bibinfo{person}{Luka Murn}, \bibinfo{person}{Marta Mrak}, {and} \bibinfo{person}{Marcel Worring}.} \bibinfo{year}{2023}\natexlab{b}.
\newblock \showarticletitle{Query-Based Video Summarization with Pseudo Label Supervision}. In \bibinfo{booktitle}{\emph{2023 IEEE International Conference on Image Processing (ICIP)}}. IEEE, \bibinfo{pages}{1430--1434}.
\newblock


\bibitem[\protect\citeauthoryear{Huang, Shen, Zhu, Rudinac, and Kanoulas}{Huang et~al\mbox{.}}{2025a}]%
        {huang2025gradient}
\bibfield{author}{\bibinfo{person}{Jia-Hong Huang}, \bibinfo{person}{Yixian Shen}, \bibinfo{person}{Hongyi Zhu}, \bibinfo{person}{Stevan Rudinac}, {and} \bibinfo{person}{Evangelos Kanoulas}.} \bibinfo{year}{2025}\natexlab{a}.
\newblock \showarticletitle{Gradient weight-normalized low-rank projection for efficient llm training}. In \bibinfo{booktitle}{\emph{Proceedings of the AAAI Conference on Artificial Intelligence}}, Vol.~\bibinfo{volume}{39}. \bibinfo{pages}{24123--24131}.
\newblock


\bibitem[\protect\citeauthoryear{Huang and Worring}{Huang and Worring}{2020}]%
        {huang2020query}
\bibfield{author}{\bibinfo{person}{Jia-Hong Huang} {and} \bibinfo{person}{Marcel Worring}.} \bibinfo{year}{2020}\natexlab{}.
\newblock \showarticletitle{Query-controllable video summarization}. In \bibinfo{booktitle}{\emph{Proceedings of the International Conference on Multimedia Retrieval}}. \bibinfo{pages}{242--250}.
\newblock


\bibitem[\protect\citeauthoryear{Huang, Wu, and Worring}{Huang et~al\mbox{.}}{2021b}]%
        {huang2021contextualized}
\bibfield{author}{\bibinfo{person}{Jia-Hong Huang}, \bibinfo{person}{Ting-Wei Wu}, {and} \bibinfo{person}{Marcel Worring}.} \bibinfo{year}{2021}\natexlab{b}.
\newblock \showarticletitle{Contextualized keyword representations for multi-modal retinal image captioning}. In \bibinfo{booktitle}{\emph{Proceedings of the 2021 International Conference on Multimedia Retrieval}}. \bibinfo{pages}{645--652}.
\newblock


\bibitem[\protect\citeauthoryear{Huang, Wu, Yang, Shi, Lin, Tegner, Worring, et~al\mbox{.}}{Huang et~al\mbox{.}}{2022a}]%
        {huang2022non}
\bibfield{author}{\bibinfo{person}{Jia-Hong Huang}, \bibinfo{person}{Ting-Wei Wu}, \bibinfo{person}{C-H~Huck Yang}, \bibinfo{person}{Zenglin Shi}, \bibinfo{person}{I Lin}, \bibinfo{person}{Jesper Tegner}, \bibinfo{person}{Marcel Worring}, {et~al\mbox{.}}} \bibinfo{year}{2022}\natexlab{a}.
\newblock \showarticletitle{Non-local attention improves description generation for retinal images}. In \bibinfo{booktitle}{\emph{Proceedings of the IEEE/CVF winter conference on applications of computer vision}}. \bibinfo{pages}{1606--1615}.
\newblock


\bibitem[\protect\citeauthoryear{Huang, Wu, Yang, and Worring}{Huang et~al\mbox{.}}{2021c}]%
        {huang2021deep}
\bibfield{author}{\bibinfo{person}{Jia-Hong Huang}, \bibinfo{person}{Ting-Wei Wu}, \bibinfo{person}{Chao-Han~Huck Yang}, {and} \bibinfo{person}{Marcel Worring}.} \bibinfo{year}{2021}\natexlab{c}.
\newblock \showarticletitle{Deep context-encoding network for retinal image captioning}. In \bibinfo{booktitle}{\emph{2021 IEEE International Conference on Image Processing (ICIP)}}. IEEE, \bibinfo{pages}{3762--3766}.
\newblock


\bibitem[\protect\citeauthoryear{Huang, Wu, Yang, and Worring}{Huang et~al\mbox{.}}{2021d}]%
        {huang2021longer}
\bibfield{author}{\bibinfo{person}{Jia-Hong Huang}, \bibinfo{person}{Ting-Wei Wu}, \bibinfo{person}{Chao-Han~Huck Yang}, {and} \bibinfo{person}{Marcel Worring}.} \bibinfo{year}{2021}\natexlab{d}.
\newblock \showarticletitle{Longer Version for "Deep Context-Encoding Network for Retinal Image Captioning"}.
\newblock \bibinfo{journal}{\emph{arXiv preprint arXiv:2105.14538}} (\bibinfo{year}{2021}).
\newblock


\bibitem[\protect\citeauthoryear{Huang, Yang, Shen, Pacces, and Kanoulas}{Huang et~al\mbox{.}}{2024b}]%
        {huang2024optimizing}
\bibfield{author}{\bibinfo{person}{Jia-Hong Huang}, \bibinfo{person}{Chao-Chun Yang}, \bibinfo{person}{Yixian Shen}, \bibinfo{person}{Alessio~M Pacces}, {and} \bibinfo{person}{Evangelos Kanoulas}.} \bibinfo{year}{2024}\natexlab{b}.
\newblock \showarticletitle{Optimizing Numerical Estimation and Operational Efficiency in the Legal Domain through Large Language Models}. In \bibinfo{booktitle}{\emph{ACM International Conference on Information and Knowledge Management (CIKM)}}.
\newblock


\bibitem[\protect\citeauthoryear{Huang, Yang, Chen, Brown, and Worring}{Huang et~al\mbox{.}}{2022b}]%
        {huang2022causal}
\bibfield{author}{\bibinfo{person}{Jia-Hong Huang}, \bibinfo{person}{Chao-Han~Huck Yang}, \bibinfo{person}{Pin-Yu Chen}, \bibinfo{person}{Andrew Brown}, {and} \bibinfo{person}{Marcel Worring}.} \bibinfo{year}{2022}\natexlab{b}.
\newblock \showarticletitle{Causal video summarizer for video exploration}. In \bibinfo{booktitle}{\emph{2022 IEEE International Conference on Multimedia and Expo (ICME)}}. IEEE, \bibinfo{pages}{1--6}.
\newblock


\bibitem[\protect\citeauthoryear{Huang, Yang, Chen, Chen, and Worring}{Huang et~al\mbox{.}}{2023c}]%
        {huang2023causalainer}
\bibfield{author}{\bibinfo{person}{Jia-Hong Huang}, \bibinfo{person}{Chao-Han~Huck Yang}, \bibinfo{person}{Pin-Yu Chen}, \bibinfo{person}{Min-Hung Chen}, {and} \bibinfo{person}{Marcel Worring}.} \bibinfo{year}{2023}\natexlab{c}.
\newblock \showarticletitle{Causalainer: Causal Explainer for Automatic Video Summarization}. In \bibinfo{booktitle}{\emph{Proceedings of the IEEE/CVF Conference on Computer Vision and Pattern Recognition}}. \bibinfo{pages}{2629--2635}.
\newblock


\bibitem[\protect\citeauthoryear{Huang, Yang, Chen, Chen, and Worring}{Huang et~al\mbox{.}}{2024a}]%
        {huang2023conditional}
\bibfield{author}{\bibinfo{person}{Jia-Hong Huang}, \bibinfo{person}{Chao-Han~Huck Yang}, \bibinfo{person}{Pin-Yu Chen}, \bibinfo{person}{Min-Hung Chen}, {and} \bibinfo{person}{Marcel Worring}.} \bibinfo{year}{2024}\natexlab{a}.
\newblock \showarticletitle{Conditional Modeling Based Automatic Video Summarization}.
\newblock \bibinfo{journal}{\emph{ACM Transactions on Multimedia Computing, Communications, and Applications (Under review)}} (\bibinfo{year}{2024}).
\newblock


\bibitem[\protect\citeauthoryear{Huang, Yang, Liu, Tian, Liu, Wu, Lin, Wang, Morikawa, Chang, et~al\mbox{.}}{Huang et~al\mbox{.}}{2021e}]%
        {huang2021deepopht}
\bibfield{author}{\bibinfo{person}{Jia-Hong Huang}, \bibinfo{person}{C-H~Huck Yang}, \bibinfo{person}{Fangyu Liu}, \bibinfo{person}{Meng Tian}, \bibinfo{person}{Yi-Chieh Liu}, \bibinfo{person}{Ting-Wei Wu}, \bibinfo{person}{I Lin}, \bibinfo{person}{Kang Wang}, \bibinfo{person}{Hiromasa Morikawa}, \bibinfo{person}{Hernghua Chang}, {et~al\mbox{.}}} \bibinfo{year}{2021}\natexlab{e}.
\newblock \showarticletitle{DeepOpht: medical report generation for retinal images via deep models and visual explanation}. In \bibinfo{booktitle}{\emph{Proceedings of the IEEE/CVF winter conference on applications of computer vision}}. \bibinfo{pages}{2442--2452}.
\newblock


\bibitem[\protect\citeauthoryear{Huang, Zhu, Shen, Rudinac, and Kanoulas}{Huang et~al\mbox{.}}{2025b}]%
        {huang2025image2text2image}
\bibfield{author}{\bibinfo{person}{Jia-Hong Huang}, \bibinfo{person}{Hongyi Zhu}, \bibinfo{person}{Yixian Shen}, \bibinfo{person}{Stevan Rudinac}, {and} \bibinfo{person}{Evangelos Kanoulas}.} \bibinfo{year}{2025}\natexlab{b}.
\newblock \showarticletitle{Image2Text2Image: A Novel Framework for Label-Free Evaluation of Image-to-Text Generation with Text-to-Image Diffusion Models}. In \bibinfo{booktitle}{\emph{International Conference on Multimedia Modeling (MMM)}}.
\newblock


\bibitem[\protect\citeauthoryear{Huang, Zhu, Shen, Rudinac, Pacces, and Kanoulas}{Huang et~al\mbox{.}}{2024c}]%
        {huang2024novel}
\bibfield{author}{\bibinfo{person}{Jia-Hong Huang}, \bibinfo{person}{Hongyi Zhu}, \bibinfo{person}{Yixian Shen}, \bibinfo{person}{Stevan Rudinac}, \bibinfo{person}{Alessio~M. Pacces}, {and} \bibinfo{person}{Evangelos Kanoulas}.} \bibinfo{year}{2024}\natexlab{c}.
\newblock \showarticletitle{A Novel Evaluation Framework for Image2Text Generation}. In \bibinfo{booktitle}{\emph{International ACM SIGIR Conference on Research and Development in Information Retrieval, LLM4Eval Workshop}}.
\newblock


\bibitem[\protect\citeauthoryear{Huck~Yang, Liu, Huang, Tian, I-Hung~Lin, Liu, Morikawa, Yang, and Tegner}{Huck~Yang et~al\mbox{.}}{2019}]%
        {huck2019auto}
\bibfield{author}{\bibinfo{person}{C-H Huck~Yang}, \bibinfo{person}{Fangyu Liu}, \bibinfo{person}{Jia-Hong Huang}, \bibinfo{person}{Meng Tian}, \bibinfo{person}{MD I-Hung~Lin}, \bibinfo{person}{Yi~Chieh Liu}, \bibinfo{person}{Hiromasa Morikawa}, \bibinfo{person}{Hao-Hsiang Yang}, {and} \bibinfo{person}{Jesper Tegner}.} \bibinfo{year}{2019}\natexlab{}.
\newblock \showarticletitle{Auto-classification of retinal diseases in the limit of sparse data using a two-streams machine learning model}. In \bibinfo{booktitle}{\emph{Computer Vision--ACCV 2018 Workshops: 14th Asian Conference on Computer Vision, Perth, Australia, December 2--6, 2018, Revised Selected Papers 14}}. Springer, \bibinfo{pages}{323--338}.
\newblock


\bibitem[\protect\citeauthoryear{Jing, Xie, and Xing}{Jing et~al\mbox{.}}{2018}]%
        {jing-etal-2018-automatic}
\bibfield{author}{\bibinfo{person}{Baoyu Jing}, \bibinfo{person}{Pengtao Xie}, {and} \bibinfo{person}{Eric Xing}.} \bibinfo{year}{2018}\natexlab{}.
\newblock \showarticletitle{On the Automatic Generation of Medical Imaging Reports}. In \bibinfo{booktitle}{\emph{ACL}}. \bibinfo{publisher}{ACL}, \bibinfo{address}{Melbourne, Australia}, \bibinfo{pages}{2577--2586}.
\newblock


\bibitem[\protect\citeauthoryear{Liu, Liu, and Sun}{Liu et~al\mbox{.}}{2019a}]%
        {liu2019aligning}
\bibfield{author}{\bibinfo{person}{Fenglin Liu}, \bibinfo{person}{Yuanxin Liu}, {and} \bibinfo{person}{Xu Sun}.} \bibinfo{year}{2019}\natexlab{a}.
\newblock \bibinfo{title}{Aligning Visual Regions and Textual Concepts for Semantic-Grounded Image Representations}.
\newblock
\newblock


\bibitem[\protect\citeauthoryear{Liu, Yang, Huck~Yang, Huang, Tian, Morikawa, Tsai, and Tegner}{Liu et~al\mbox{.}}{2019b}]%
        {liu2019synthesizing}
\bibfield{author}{\bibinfo{person}{Yi-Chieh Liu}, \bibinfo{person}{Hao-Hsiang Yang}, \bibinfo{person}{C-H Huck~Yang}, \bibinfo{person}{Jia-Hong Huang}, \bibinfo{person}{Meng Tian}, \bibinfo{person}{Hiromasa Morikawa}, \bibinfo{person}{Yi-Chang~James Tsai}, {and} \bibinfo{person}{Jesper Tegner}.} \bibinfo{year}{2019}\natexlab{b}.
\newblock \showarticletitle{Synthesizing new retinal symptom images by multiple generative models}. In \bibinfo{booktitle}{\emph{Computer Vision--ACCV 2018 Workshops: 14th Asian Conference on Computer Vision, Perth, Australia, December 2--6, 2018, Revised Selected Papers 14}}. Springer, \bibinfo{pages}{235--250}.
\newblock


\bibitem[\protect\citeauthoryear{Lu, Yang, Batra, and Parikh}{Lu et~al\mbox{.}}{2017}]%
        {lu2017hierarchical}
\bibfield{author}{\bibinfo{person}{Jiasen Lu}, \bibinfo{person}{Jianwei Yang}, \bibinfo{person}{Dhruv Batra}, {and} \bibinfo{person}{Devi Parikh}.} \bibinfo{year}{2017}\natexlab{}.
\newblock \bibinfo{title}{Hierarchical Question-Image Co-Attention for Visual Question Answering}.
\newblock
\newblock


\bibitem[\protect\citeauthoryear{Shen, Bi, Huang, Zhu, and Anuj}{Shen et~al\mbox{.}}{2024}]%
        {shen2024parameterefficient}
\bibfield{author}{\bibinfo{person}{Yixian Shen}, \bibinfo{person}{Qi Bi}, \bibinfo{person}{Jia-Hong Huang}, \bibinfo{person}{Hongyi Zhu}, {and} \bibinfo{person}{Pathania Anuj}.} \bibinfo{year}{2024}\natexlab{}.
\newblock \showarticletitle{Parameter-Efficient Fine-Tuning via Selective Discrete Cosine Transform}.
\newblock \bibinfo{journal}{\emph{arXiv preprint arXiv:2405.13372}} (\bibinfo{year}{2024}).
\newblock


\bibitem[\protect\citeauthoryear{Shen, Bi, Huang, Zhu, Pimentel, and Pathania}{Shen et~al\mbox{.}}{2025a}]%
        {shen2025macp}
\bibfield{author}{\bibinfo{person}{Yixian Shen}, \bibinfo{person}{Qi Bi}, \bibinfo{person}{Jia-Hong Huang}, \bibinfo{person}{Hongyi Zhu}, \bibinfo{person}{Andy~D Pimentel}, {and} \bibinfo{person}{Anuj Pathania}.} \bibinfo{year}{2025}\natexlab{a}.
\newblock \showarticletitle{MaCP: Minimal yet Mighty Adaptation via Hierarchical Cosine Projection}. In \bibinfo{booktitle}{\emph{Proceedings of the 63nd Annual Meeting of the Association for Computational Linguistics (Long Papers)}}.
\newblock


\bibitem[\protect\citeauthoryear{Shen, Bi, Huang, Zhu, Pimentel, and Pathania}{Shen et~al\mbox{.}}{2025b}]%
        {shen2025ssh}
\bibfield{author}{\bibinfo{person}{Yixian Shen}, \bibinfo{person}{Qi Bi}, \bibinfo{person}{Jia-Hong Huang}, \bibinfo{person}{Hongyi Zhu}, \bibinfo{person}{Andy~D Pimentel}, {and} \bibinfo{person}{Anuj Pathania}.} \bibinfo{year}{2025}\natexlab{b}.
\newblock \showarticletitle{Ssh: Sparse spectrum adaptation via discrete hartley transformation}. In \bibinfo{booktitle}{\emph{Proceedings of the 2025 Conference of the North American Chapter of the Association for Computational Linguistics: Human Language Technologies}}.
\newblock


\bibitem[\protect\citeauthoryear{Tukey and Wiener}{Tukey and Wiener}{2014}]%
        {tukey2014impact}
\bibfield{author}{\bibinfo{person}{Melissa~H Tukey} {and} \bibinfo{person}{Renda~Soylemez Wiener}.} \bibinfo{year}{2014}\natexlab{}.
\newblock \showarticletitle{The impact of a medical service on patient safety, quality and resident training}.
\newblock \bibinfo{journal}{\emph{JGIM}} (\bibinfo{year}{2014}).
\newblock


\bibitem[\protect\citeauthoryear{Wang, Yang, Bartz, and Meinel}{Wang et~al\mbox{.}}{2016}]%
        {wang2016image}
\bibfield{author}{\bibinfo{person}{Cheng Wang}, \bibinfo{person}{Haojin Yang}, \bibinfo{person}{Christian Bartz}, {and} \bibinfo{person}{Christoph Meinel}.} \bibinfo{year}{2016}\natexlab{}.
\newblock \showarticletitle{Image captioning with deep bidirectional LSTMs}. In \bibinfo{booktitle}{\emph{ACM MM}}. ACM, \bibinfo{pages}{988--997}.
\newblock


\bibitem[\protect\citeauthoryear{Wang, Zhang, Huang, Rudinac, Kackovic, Wijnberg, and Worring}{Wang et~al\mbox{.}}{2024}]%
        {wang2024ada}
\bibfield{author}{\bibinfo{person}{Shuai Wang}, \bibinfo{person}{David~W Zhang}, \bibinfo{person}{Jia-Hong Huang}, \bibinfo{person}{Stevan Rudinac}, \bibinfo{person}{Monika Kackovic}, \bibinfo{person}{Nachoem Wijnberg}, {and} \bibinfo{person}{Marcel Worring}.} \bibinfo{year}{2024}\natexlab{}.
\newblock \showarticletitle{Ada-HGNN: Adaptive Sampling for Scalable Hypergraph Neural Networks}.
\newblock \bibinfo{journal}{\emph{arXiv preprint arXiv:2405.13372}} (\bibinfo{year}{2024}).
\newblock


\bibitem[\protect\citeauthoryear{Wu, Huang, Lin, and Worring}{Wu et~al\mbox{.}}{2023}]%
        {wu2023expert}
\bibfield{author}{\bibinfo{person}{Ting-Wei Wu}, \bibinfo{person}{Jia-Hong Huang}, \bibinfo{person}{Joseph Lin}, {and} \bibinfo{person}{Marcel Worring}.} \bibinfo{year}{2023}\natexlab{}.
\newblock \showarticletitle{Expert-defined Keywords Improve Interpretability of Retinal Image Captioning}. In \bibinfo{booktitle}{\emph{Proceedings of the IEEE/CVF Winter Conference on Applications of Computer Vision}}. \bibinfo{pages}{1859--1868}.
\newblock


\bibitem[\protect\citeauthoryear{Xu, Ba, Kiros, Cho, Courville, Salakhutdinov, Zemel, and Bengio}{Xu et~al\mbox{.}}{2016}]%
        {xu2016show}
\bibfield{author}{\bibinfo{person}{Kelvin Xu}, \bibinfo{person}{Jimmy Ba}, \bibinfo{person}{Ryan Kiros}, \bibinfo{person}{Kyunghyun Cho}, \bibinfo{person}{Aaron Courville}, \bibinfo{person}{Ruslan Salakhutdinov}, \bibinfo{person}{Richard Zemel}, {and} \bibinfo{person}{Yoshua Bengio}.} \bibinfo{year}{2016}\natexlab{}.
\newblock \bibinfo{title}{Show, Attend and Tell: Neural Image Caption Generation with Visual Attention}.
\newblock
\newblock
\showeprint[arxiv]{cs.LG/1502.03044}


\bibitem[\protect\citeauthoryear{Yang, Huang, Liu, Chiu, Gao, Lyu, Tegner, et~al\mbox{.}}{Yang et~al\mbox{.}}{2018}]%
        {yang2018novel}
\bibfield{author}{\bibinfo{person}{C-H~Huck Yang}, \bibinfo{person}{Jia-Hong Huang}, \bibinfo{person}{Fangyu Liu}, \bibinfo{person}{Fang-Yi Chiu}, \bibinfo{person}{Mengya Gao}, \bibinfo{person}{Weifeng Lyu}, \bibinfo{person}{Jesper Tegner}, {et~al\mbox{.}}} \bibinfo{year}{2018}\natexlab{}.
\newblock \showarticletitle{A novel hybrid machine learning model for auto-classification of retinal diseases}.
\newblock \bibinfo{journal}{\emph{Joint International Conference on Machine Learning (ICML) and International Joint Conference on Artificial Intelligence (IJCAI) Workshop on Computational Biology}} (\bibinfo{year}{2018}).
\newblock


\bibitem[\protect\citeauthoryear{You, Jin, Wang, Fang, and Luo}{You et~al\mbox{.}}{2016}]%
        {you2016image}
\bibfield{author}{\bibinfo{person}{Quanzeng You}, \bibinfo{person}{Hailin Jin}, \bibinfo{person}{Zhaowen Wang}, \bibinfo{person}{Chen Fang}, {and} \bibinfo{person}{Jiebo Luo}.} \bibinfo{year}{2016}\natexlab{}.
\newblock \bibinfo{title}{Image Captioning with Semantic Attention}.
\newblock
\newblock
\showeprint[arxiv]{cs.CV/1603.03925}


\bibitem[\protect\citeauthoryear{Zhang, Aliannejadi, Pei, Yuan, Huang, and Kanoulas}{Zhang et~al\mbox{.}}{2024a}]%
        {zhang2024comparative}
\bibfield{author}{\bibinfo{person}{Weijia Zhang}, \bibinfo{person}{Mohammad Aliannejadi}, \bibinfo{person}{Jiahuan Pei}, \bibinfo{person}{Yifei Yuan}, \bibinfo{person}{Jia-Hong Huang}, {and} \bibinfo{person}{Evangelos Kanoulas}.} \bibinfo{year}{2024}\natexlab{a}.
\newblock \showarticletitle{A Comparative Analysis of Faithfulness Metrics and Humans in Citation Evaluation}. In \bibinfo{booktitle}{\emph{International ACM SIGIR Conference on Research and Development in Information Retrieval, LLM4Eval Workshop}}.
\newblock


\bibitem[\protect\citeauthoryear{Zhang, Aliannejadi, Yuan, Pei, Huang, and Kanoulas}{Zhang et~al\mbox{.}}{2024b}]%
        {zhang2024towards}
\bibfield{author}{\bibinfo{person}{Weijia Zhang}, \bibinfo{person}{Mohammad Aliannejadi}, \bibinfo{person}{Yifei Yuan}, \bibinfo{person}{Jiahuan Pei}, \bibinfo{person}{Jia-Hong Huang}, {and} \bibinfo{person}{Evangelos Kanoulas}.} \bibinfo{year}{2024}\natexlab{b}.
\newblock \showarticletitle{Towards Fine-Grained Citation Evaluation in Generated Text: A Comparative Analysis of Faithfulness Metrics}. In \bibinfo{booktitle}{\emph{Proceedings of the 2024 International Natural Language Generation Conference (INLG)}}.
\newblock


\bibitem[\protect\citeauthoryear{Zhang, Huang, Vakulenko, Xu, Rajapakse, and Kanoulas}{Zhang et~al\mbox{.}}{2024c}]%
        {zhang2024beyond}
\bibfield{author}{\bibinfo{person}{Weijia Zhang}, \bibinfo{person}{Jia-Hong Huang}, \bibinfo{person}{Svitlana Vakulenko}, \bibinfo{person}{Yumo Xu}, \bibinfo{person}{Thilina Rajapakse}, {and} \bibinfo{person}{Evangelos Kanoulas}.} \bibinfo{year}{2024}\natexlab{c}.
\newblock \showarticletitle{Beyond Relevant Documents: A Knowledge-Intensive Approach for Query-Focused Summarization using Large Language Models}. In \bibinfo{booktitle}{\emph{Proceedings of the 2024 International Conference on Pattern Recognition (ICPR)}}.
\newblock


\bibitem[\protect\citeauthoryear{Zhang, Pal, Huang, Kanoulas, and de~Rijke}{Zhang et~al\mbox{.}}{2024d}]%
        {zhang2024qfmts}
\bibfield{author}{\bibinfo{person}{Weijia Zhang}, \bibinfo{person}{Vaishali Pal}, \bibinfo{person}{Jia-Hong Huang}, \bibinfo{person}{Evangelos Kanoulas}, {and} \bibinfo{person}{Maarten de Rijke}.} \bibinfo{year}{2024}\natexlab{d}.
\newblock \showarticletitle{QFMTS: Generating Query-Focused Summaries over Multi-Table Inputs}. In \bibinfo{booktitle}{\emph{Proceedings of the 2024 European Conference on Artificial Intelligence (ECAI)}}.
\newblock


\bibitem[\protect\citeauthoryear{Zhu, Huang, Rudinac, and Kanoulas}{Zhu et~al\mbox{.}}{2024}]%
        {zhu2024enhancing}
\bibfield{author}{\bibinfo{person}{Hongyi Zhu}, \bibinfo{person}{Jia-Hong Huang}, \bibinfo{person}{Stevan Rudinac}, {and} \bibinfo{person}{Evangelos Kanoulas}.} \bibinfo{year}{2024}\natexlab{}.
\newblock \showarticletitle{Enhancing Interactive Image Retrieval With Query Rewriting Using Large Language Models and Vision Language Models}. In \bibinfo{booktitle}{\emph{Proceedings of the 2024 International Conference on Multimedia Retrieval}}. \bibinfo{pages}{978--987}.
\newblock


\end{thebibliography}

\end{document}